# Model for unidirectional movement of axonemal and cytoplasmic dynein molecules

Ping Xie*, Shuo-Xing Dou, and Peng-Ye Wang

*Laboratory of Soft Matter Physics, Beijing National Laboratory for Condensed Matter Physics, Institute of Physics, Chinese Academy of Sciences, Beijing 100080, China*

**Abstract**

A model for the unidirectional movement of dynein is presented based on structural observations and biochemical experimental results available. In this model, the binding affinity of dynein for microtubule is independent of its nucleotide state and the change between strong and weak microtubule-binding is determined naturally by the variation of relative orientation between the stalk and microtubule as the stalk rotates following nucleotide-state transition. Thus the enigmatic communication from the ATP binding site in the globular domain to the far MT-binding site in the tip of the stalk, which is prerequisite in conventional models, is not required. Using the present model, the previous experimental results such as the effect of ATP and ADP bindings on dissociation of dynein from microtubule, the processive movement of single-headed axonemal dyneins at saturating ATP concentration, the load dependence of step size for the processive movement of two-headed cytoplasmic dyneins and the dependence of stall force on ATP concentration can be well explained.

\* Corresponding author.
*E-mail address*: pxie@aphy.iphy.ac.cn

# 1. INTRODUCTION

Dyneins are microtubule (MT) based motor proteins, which fall broadly into two principle classes: axonemal dynein and cytoplasmic dynein [1–3]. Axonemal dynein that was the first to be discovered functions as a molecular engine for ciliary and flagellar movement [4]. More than 20 years after the first discovery of the axonemal dynein, cytoplasmic dynein was identified [5] and found to be involved in transport of organelles and vesicles as well as in spindle assembly and chromosome segregation [6,7]. Each dynein is a complex of 1 to 3 heavy chains, each with a relative molecular mass greater than 500,000, together with a number of intermediate and light chains. Each heavy chain constitutes the fundamental motor unit. Electron microscopy has established that the heavy chain folds to form a globular domain with two elongated structures, the stalk and stem, emerging from it. The stalk and the stem bind the MT track and cargo, respectively. The stalk is most probably an anti-parallel coiled-coil, with a small MT-binding domain at its tip.

With extensive investigations using different experimental methods, such as biochemical, biophysical, and single-molecular approaches, many dynamical behaviors of dyneins *in vitro* have been elucidated. Important mechanical properties such as stall force, step size and velocity have been determined [8–12]. However, the microscopic mechanism of its movement is still not very clear. Generally, two types of models have been proposed. One type is the thermal ratchet model in which the motor is simply viewed as a Brownian particle moving in two (or more) periodic but spatially asymmetric stochastically switched potentials [13,14]. Another prevailing one is the power-stroke model that is proposed based on structural observations [12,15,16], with which some dynamical behaviours are quantitatively simulated numerically [17,18]. In this model, the power stroke is generated by the relative rotation between the stalk and stem when dynein makes a transition from ADP.Vi-state to apo-state (or nucleotide-free state). In order to have a unidirectional movement, it has been assumed that ATP binding to the globular domain reduces the MT-binding affinity. That means that a communication should exist from the ATP-binding site in the globular domain to the far MT-binding site in the tip of the stalk, which, however, is difficult to imagine in view of the dynein structure.

In this work we present another power-stroke model for the unidirectional movement of dynein based on available structural observations. Our model is different from the conventional models in that, in our model, the change between strong and weak MT-bindings of the stalk tip following nucleotide-state transitions does not rely on a communication from the ATP to MT binding sites, but is resulted naturally from the varying orientation of the stalk with respect to MT during the stalk rotation. As will be shown in the following Sections 2–4, the results deduced from this model not only show agreement with available single-molecule experimental results such as the processive movement of single-headed axonemal dyneins at saturating ATP concentration [9], the load dependence of step size for the processive movement of two-headed cytoplasmic dyneins and the dependence of stall force on ATP concentration [12], but also are consistent with previous available biochemical experimental results such as the nucleotide-binding-dependent dissociation of dyneins from MT [19–24].

## 2. MODEL

In this work, we propose our model based on the following two points.

**(i)** *There are two nucleotide-dependent conformations of dynein.* The conformation of dynein in ADP.Vi state is schematically shown in Fig. 1*a* and that in apo-state shown in Fig. 1*b*. In other words, the release of ATP hydrolysis products leads to the rotation of the stalk (or stem), i.e., the change in the relative orientation between the stalk and the stem, from that as shown in Fig. 1*a* to that as shown in Fig. 1*b*, and ATP binding has the opposite effect, i.e., leading to the change in the relative orientation from that as shown in Fig. 1*b* to that as shown in Fig. 1*a*.

This point is supported by structural study of the inner-arm dynein c of *Chlamydomonas* flagella using electron microscopy [16]. The origin of the nucleotide-dependent conformational change is not clear. We present one possible origin as follows. It has been experimentally demonstrated that, of the six AAA modules, only AAA1-AAA4 modules have ATP-binding loops and only the AAA1 (P1) and AAA3 (P3) loops may have ATPase activities [25]. Considering that the ATP-binding/hydrolysis loop lies in the interfaces between AAA1 and AAA2 (and/or between AAA3 and AAA4) [25], it is reasonable for us to assume that ATP binding tightens the contacts between AAA1 and AAA2 (and/or between AAA3 and AAA4) whereas the release of hydrolysis products loosens the contacts. In other words, ATP binding leads to tighter conformations of AAA1 and AAA2 (and/or of AAA3 and AAA4). As schematically shown in Fig. 1, this tightening or loosening induces a change in the relative orientation between the stalk and the stem. Note that this characteristic is much similar to other AAA-domain proteins such as F1-ATPase motor, where the nucleotide-free state also corresponds to a loose conformation while nucleotide-binding state to a tight conformation [26].

The nucleotide-dependent tightening/loosing of the contacts as shown in Fig. 1 is in agreement with the recent FRET experimental results by Kon et al. [27], where it was reported that the dynein adopts two conformational states (states I and II) in the course of its ATP hydrolysis cycle: In state II the stem is proximal to AAA2 module while in state I the stem is away from AAA2 module. ATP binding induced the stem motion from state I to state II, i.e., the stem becoming close to AAA2 module (Fig. 1a), while ADP-releasing induces the stem motion from state II to state I, i.e., the stem becoming away from AAA2 module (Fig. 1b). The nucleotide-dependent compact of the ring as shown in Fig. 1 is also consistent with the proposal by Höök et al. [28]: ATP-binding induces a compact conformation of the ring (Fig. 1a) while ADP-releasing loosens the ring (Fig. 1b).

Another possible origin of the nucleotide-dependent conformational change was proposed by Burgess et al. [16], where the orientation of AAA1-AAA4 modules relative to each other might change as a function of nucleotide binding or release. These change help to drive the movement of the globular domain relative to the stem attached at the AAA1 module (see Fig. 2). The origin for the change in the relative orientation between the stalk and the stem may be one or both of the above two cases.

**(ii)** *Strong MT-binding of the stalk tip promotes a conformational change in the active site, thus activating release of the ATP hydrolysis product ADP.* This point is supported by the experimental observation of MT-activated ATPase activity [21,22,24,28,29]. For example, it was shown that MT can enhance greatly the product-release rate by more than ten times though it has negligible effect on ATP-binding rate for *Tetrahymena* 22S dynein [29]. We will discuss this MT-activated product release in Section 4 (Discussion).

**2.1. Axonemal dyneins**

Based on the above two points we propose a model for the unidirectional movement of MT by a monomeric axonemal dynein motor, as observed in *in vitro* MT sliding study by Sakakibara et al. [9], where the dynein is fixed and MT can be moved freely. As the binding affinity of the stalk tip for MT is independent of nucleotide states, the dynein may bind strongly to MT in either of the two nucleotide states (conformational states), i.e., ADP.Vi-state and apo-state. (As we will discuss in detail in Section 4, this argument of nucleotide-*independent* binding affinity is consistent with previous available biochemical experimental results including the nucleotide-binding-*dependent* dissociation of dyneins from MT [19–24].) We thus consider the two cases separately.

**(i)** *Dynein binds strongly to MT in ADP.Vi state at the beginning*

In this case (Fig. 3*a*), the Pi and ADP are released rapidly [see Point (ii)]. This leads to the rotation of the stalk [see Point (i)], thus driving MT toward the plus end (Fig. 3*b*). Since the stalk can be considered nearly as rigid [30–32], it is evident that, as the stalk rotates, the contacting surface between the stalk tip and MT is reduced and thus the MT-binding strength of the stalk tip decreases. When the binding force, $F_b$, becomes smaller than the Stokes force, $F_S$, acted on the MT, the unidirectional movement of MT ceases although the stalk tip may still move. Here $F_S = \Gamma v = \Gamma L \omega$, with $\Gamma$ being the drag coefficient on MT, $L$ the stalk length and $\omega$ the rotation velocity of the stalk.

Then upon ATP binding, the stalk rotates back and the dynein returns to its original conformational state [see Point (i)]. After dynein binds strongly to MT again (Fig. 3*c*), an ATPase cycle is finished, with the MT on average moving a distance of *nd*, where *d* = 8 nm is the period of the MT lattice and *n* is an integer. The integer *n* is determined by $L_S$ through the following inequality, $(n-1/2)d < L_S < (n+1/2)d$, where $L_S$ is the moving distance of MT from Fig. 3*a* to *b*. Electron microscopy observation [16] suggested that the mean displacement of the stalk tip is about $d_{stalk}$ = 13~15 nm. This implies that $L_S$ should be smaller than 13~15 nm. If $L_S < 12$ nm one has *n* = 1, which explains why in most experiments only 8-nm step size for axonemal dynein was reported [9,33].

It should be mentioned here that in the above discussion we have regarded both the stem and stalk as being rigid. Previous results showed that the stalk can be considered nearly as rigid [30–32]. The stem, however, may have some flexibility (or elasticity). To consider the effect of elasticity of the stem, we still start with the configuration as shown in Fig. 3a but assume that MT is fixed. After release of the products Pi and ADP the relative rotation between the stem and stalk would result in a conformational change to that as shown in Fig. 4. In this configuration (Fig. 4), due to bending of the stem, there exists an internal elastic force. Now if MT becomes free the internal elastic force will induce the system to a state that is the same as that as shown in Fig. 3b, in which no internal elastic force exists. Thus whether the stem is rigid or behaves elastically (except very flexible) the moving distance of MT should be nearly the same. If the stem is very flexible, the relative orientation between the stalk and stem would be unable to exert even a very small force and thus the present as well as the conventional power-stroke models could not work. Therefore, we consider that the stem is not very flexible but should have a finite bending elasticity.

It is interesting to note that, at saturating ATP as in the experiment by Sakakibara et al. [9], after the stalk rotation (Fig. 3*b*) the stalk will return back immediately and then bind strongly to MT again (Fig. 3*c*). In other words, the time spent in the weak binding state (Fig. 3*b*) of the single-headed dynein is sufficiently short, so that the stalk is able to rebind strongly to MT (Fig. 3*c*) before MT is diffused away by thermal noise. In such a way, a single monomeric dynein can drive MT to move processively, which well explains the experimental results [9]. However, according to the conventional models, both ATP and ADP.Pi states are weak MT-binding states. Thus the total time spent in the weak binding states is not short and a monomeric dynein *could not* move MT *processively*.

From above discussion it is noted that, in order for a single-headed dynein to drive the processive movement of MT at saturating ATP, the dynein should spend most of its enzymatic cycle in a strong binding state, which means a high duty ratio. However, the reported experimental data [9] show that the velocity of MT translation depended strongly on motor surface density, meaning a low duty ratio. This contradiction can be explained from our model as follows: When there are many dyneins bound to MT, one dynein may change from strong binding (Fig. 3*a*) to weak binding even if its conformational (nucleotide) state remains unchanged due to the translation of MT driven by other dyneins. This is equivalent to a decrease of the strong-binding time or a decrease of the duty ratio of one dynein. This behavior is in contrast with other motors such as kinesin and myosin, the duty ratio (or ratio of strong-binding time to total ATPase time) of which is only determined by the nucleotide states [34].

(**ii**) *Dynein binds strongly to MT in apo-state at the beginning*

After ATP binding to dynein in the apo-state (Fig. 3*a'*), the stalk rotates from the apo-state orientation to the ADP.Vi-state orientation (Fig. 3*b'*). This drives the MT moving toward the minus end. As discussed in Case (i), with the rotation of the stalk the interaction surface between the stalk tip and MT is reduced and thus the MT-binding strength of the stalk tip also decreases. At the end of the stalk rotation, the stalk tip is detached from MT. At present, due to the very low rate of product release of the dynein without MT activation [see Point (ii)], within the long-time period of product release, MT is most probably diffused away by thermal noise (not shown), or the stalk tip will rebind strongly to MT while driving MT toward the plus direction, as shown in Fig. 3*c'*. The conformational state in Fig. 3*c'* becomes the same as that shown in Fig. 3*a*. Thus, except for the first step (from Fig. 3*a'* to *c'*), MT will then move toward the plus end processively with a step size of *nd* (where *n* usually equals to 1) just the same as in Case (i) (Fig. 3*a-c*).

The conformational change from Fig. 3*a'* to *b'* can also explain the experimental observation that ATP binding induces detachment of dynein from MT [35]. After the ADP.Vi dynein rebinds to MT, as shown in Fig. 3*c'*, MT accelerates the release of the ATP hydrolysis products [see Point (ii)].

Experimental results demonstrated that there are two ATP hydrolysis sites (P1 loop and P3 loop) within a dynein globular domain [25]. If we adopt the assumption that ATP binding (product release) tightens (loosens) the conformation between the two adjacent AAA modules, from our model it is possible that either the ATPase cycles of the two ATP hydrolysis sites may collectively produce one power stroke or may independently produce two power strokes. We prefer to believe that an ATPase cycle of either site can produce a step. In this case, the mean displacement of the stalk tip should be half of that (~15 nm) as observed by using electron microscopy [16], i.e., about 7.5 nm, which gives $n = 1$, meaning an 8-nm step for each power stroke.

Moreover, with the assumption that ATP binding to a loop tightens the conformation between its two adjacent AAA modules, the experimental results [36] that there are two association constants for mantATP binding, with one constant being more than 10 times larger than the other one, can be explained as follows: Once an ATP binds to one loop such as P1 (P3), its conformational tightening will induce a force on the other loop P3 (P1), which results in the conformational change of the residues in vicinity of the loop P3 (P1). This may thus greatly reduce the association constant for mantATP binding of the second loop P3 (P1). Based on this idea, the experimental results by Höök et al. [28] can also be explained: The conformational loosening induced by ADP release from one loop induces a conformational change of the other loop in the full motor domain fragment. This will influence the ADP-release rate of the second loop, thus giving two different ADP-release rates with one much larger than the other one. For the half-motor domain fragment, due to the integrity of the AAA1-4 modules, the conformational loosening induced by ADP release from one loop may still induce a conformational change of the residues in vicinity of the other loop, thus also giving two different ADP-release rates [28]. It is interesting to note that the much smaller ATPase activity of the second loop resulted from ATP binding to the first loop in the intact globular domain ensures that, on average, one ATP is consumed for one power stroke.

However, it is also noted that, according to this idea, if loop P1's two adjacent modules AAA1-2 are dissected from loop P3's two adjacent modules AAA3-4, the ATPase activities of the two loops P1 and P3 should be similar. This is indeed consistent with the experimental results by Takahashi et al. [25].

If one amino acid in one of the two loops (P1 and P3) is mutated, ATP binding to the loop is blocked or greatly reduced [23]. On the other hand, the mutated amino acid in the loop may induce side-chain conformational changes of the residues in vicinity of the mutation [37]. This consequently induces a force on the other loop and thus the ATPase activity in the other loop will be also greatly reduced, similar to the effect of ATP binding to one loop on the ATPase activity of the other loop. Therefore, the experimental result that the MT-activated ATPase activity is greatly reduced when either one or both of the P1 and P3 loops is mutated [23] can be explained. The greatly reduced MT-activated ATPase activities of both P1 and P3 loops, in turn, result in rare rotation of the stalk and thus rare detachment of dynein from MT, i.e., ATP-insensitive MT binding [23,38,39]. Furthermore, the P1 mutant dynein having smaller ATPase and slower MT sliding activities than the P3 mutant [23] may be due to that the internal force resulted from the conformational change of P1 loop has a larger effect on P3 loop than the internal force on P1 loop resulted from the conformational change of P3 loop.

**2.2. Cytoplasmic dyneins**

Like conventional kinesin and myosin-V, cytoplasmic dynein is also a two-headed molecule. As adopted for homodimeric kinesin [40] and myosin-V [41], we assume that the free state of the cytoplasmic dynein homodimer is as shown in Fig. 5, with the orientations of the two heads being approximately symmetrical.

For the case that MT is fixed and dynein can move freely, as in the experiments by King and Schroer [10] and by Mallik et al. [12], we describe the processive movement of a single cytoplasmic dynein molecule along MT as follows. We begin with the two stalk tips of the dynein molecule in ADP.Vi state binding strongly to MT in fixed orientations, as shown in Fig. 6*a*. According to the moving direction, we will call the head close to the minus end of MT the leading head and that close to the plus end the trailing head. Since now the dynein dimer deviates significantly from its equilibrium conformation (Fig. 5*a*), there exists an internal elastic force and torque that act on the two heads with opposite direction. Here, we only consider saturating [ATP] and thus the ATP-binding time can be neglected. The process of ATP turnover, i.e., ATP hydrolysis and product release, is rate limiting. As the product may be released either earlier from the trailing head or earlier from the leading head, we thus consider the two cases separately.

**(i)** *Effective mechanochemical coupling*

The products are released earlier from the trailing head (Fig. 6*a*). Driven by the internal elastic force and torque, the change in the relative orientation between the stalk and the stem of the trailing head leads to its movement toward the right, as shown by dashed lines in Fig. 6*a*, thus decreasing the binding force of the stalk tip to MT. After detachment, the trailing head continues to move to its equilibrium position, as shown in Fig. 6*b* (dashed line for the stalk). After ATP binding to the new leading head its stalk rotates back (solid line for the stalk, Fig. 6*b*). Then the stalk of the new leading head will bind strongly to MT in the fixed orientation, as shown in Fig. 6*c*. Thus a mechanical step is made. The step size has a high probability of being *nd*, where *n* is an integer that approximately satisfies $(n-1/2)d < d_{equi}^{(0)} < (n+1/2)d$, with $d_{equi}^{(0)}$ being the equilibrium distance between the two stalk tips along MT (see Fig. 6*b*). It is mentioned that the step size here is mainly determined by the equilibrium distance $d_{equi}^{(0)}$ between the two stalk tips for the cytoplasmic dynein dimer, whereas the step size for an axonemal dynein monomer is mainly determined by the displacement $d_{stalk}$ of the stalk tip from one power stroke (see Sub-section 2.1). The fact that $d_{equi}^{(0)}$ is larger than $d_{stalk}$ gives that the step size for the former is larger than that for the latter.

Note that, when a backward load (opposite to moving direction) is acted on the dimer, the stem of the MT-bound trailing head can be bended elastically backwards. Thus the distance $d_{equi}$ between the two tips will decrease, as can be noted from Fig. 6*b*, which causes *n* ($d_{equi}/d - 1/2 < n < d_{equi}/d + 1/2$) to become smaller. Therefore, we have the following result: The mean step size decreases with the increase of the backward load. This is consistent with the experimental observations [12]. On the contrary, if the load acted on the dynein is directed in the forward direction, the stem of the MT-bound trailing head will be bended elastically forwards and thus the mean step size will increase. These results are different from those for conventional kinesin: Because for kinesin the equilibrium distance between its two heads $d_{equi} \approx 5$ nm [42] is smaller than $d = 8$ nm, we have $n < 1.5$. The step size for kinesin is therefore always *d*.

**(ii)** *Futile mechanochemical coupling*

The products are released earlier from the leading head (Fig. 7*a*). Driven by the internal elastic force and torque, the change in the relative orientation between the stalk and the stem of the leading head will make the dynein dimer have a conformation with a reduced internal force (Fig. 7*b*). After ATP binding to the leading head the dynein conformation is resumed, as shown in Fig. 7*c*. Thus a futile mechanochemical cycle is completed.

Therefore, we see from the above two cases that, whether the ATPase activities of the two heads are coordinated or not, the cytoplasmic dynein can move processively along MT in the unidirectional direction. If the ATPase rates of the two heads are independent and equal, on average, two ATPase cycles are coupled to one mechanical step, with one being mechanochemically effective and the other one mechanochmically futile. In fact, as we had discussed for kinesin [40] and myosin-V [41], due to the opposite internal forces acted on the two heads the ATPase rate of the trailing head can be enhanced whereas that of the leading head be reduced. Thus the ATPase rate of the trailing head would be much larger than that of the leading head, which would result in a mechanical step per ATPase cycle, i.e., a 1:1 mechanochemical coupling scenario.

## 3. DYNAMICS OF CYTOPLASMIC DYNEINS

In this section, based on the model presented in Sub-section 2.2, we study the dynamical behaviours of cytoplasmic dyneins.

### 3.1. Step-size distribution

First, we give approximate calculations of the dependence of step size on load $F_{load}$. As usually adopted in experiments, $F_{load}$ is defined as positive when it is opposite to the moving direction. Assume that the stem of the dynein has the property of a uniform and isotopic elastic beam with the bending modulus $EI$. As schematically shown in Fig. 8a, the displacement of the stem tip of the MT-bound trailing head can be calculated by using the following equation

$$\Delta = \frac{F_{load} l^3}{3EI}, \tag{1}$$

where $l$ is the length of the stem. The displacement of the stalk tip of the leading head is also approximately equal to $\Delta$.

To be consistent with the dynein structure, the stem length is about $l$ = 14 nm [3]. The persistence length of the stem is also taken as $l_p$ = 14 nm, which gives the modulus $EI = l_p k_B T \approx 57$ nm at the room temperature of 24 °C. Thus the equilibrium distance between the two stalk tips, $d_{equi} = d_{equi}^{(0)} - \Delta$, versus load $F_{load}$ can be calculated by using Eq. (1), where $d_{equi}^{(0)}$ is the equilibrium distance between the two stalk tips under no load (see Fig. 6*b*). Taking $d_{equi}^{(0)} = 23$ nm, we can obtain the distribution of step size versus $F_{load}$ by using the following Gaussian distribution,

$$P_i \propto \exp\left[-\left(d_{equi} - 8i\right)^2 / 2w^2\right], \quad (i \text{ is an integer}) \quad (2)$$

where $P_i$ is the probability for the stalk tip to bind the *i*th binding site along MT, $w$ is the half-width of the Gaussian distribution. The results are shown in Fig. 8*b*. Note that the results in Fig. 8*b* are applicable to the case of saturating [ATP].

For the case of low [ATP], the stalk tip of the leading head may be bound to MT in apo state. As can be seen from Fig. 6*b* (dashed stalk of the leading head), the equilibrium distance between the two stalk tips in this case is larger than that for the case of saturating [ATP]. This will result in the equilibrium distance $d_{equi} = d_{equi}^{(0)} - \Delta$ under load being larger than that for the case of saturating [ATP]. Taking $d_{equi}^{(0)} = 27$ nm, the calculated distribution of step size versus $F_{load}$ are shown in Fig. 8*c*. The theoretical results in Fig. 8 are consistent with the experimental ones [12]: The mean step size decreases with the increase of backward load and, furthermore, the step size is a complex admixture of 24 and 32 nm under no load and low [ATP] while is mainly 24 nm under no load and saturating [ATP].

### 3.2. Mean velocity

As we discussed in Sub-section 2.2, due to the opposite internal forces acted on the two heads of cytoplasmic dynein, the ATPase rate of the trailing head can be enhanced whereas the rate of the leading head be reduced. Thus the ATPase rate of the trailing head would be much larger than that of the leading head, which results in a mechanical step per ATPase cycle, i.e., a 1:1 mechanochemical coupling. Based on this, the mean movement velocity of the dimeric cytoplasmic dynein at saturating [ATP] can be thus written as

$$V = k_c D, \quad (3)$$

where $D$ is the mean step size calculated from the results given in Fig. 8. $k_c$ is the ATP turnover rate of the trailing head, the load dependence of which is generally written as follows [40,41,43]

$$k_c = \frac{k_c^{(0)}(1+A_c)}{1+A_c \exp(F_{trail}\delta_c/k_B T)},  \qquad (4)$$

where $F_{trail} = F_{load}/2 - F_0$ is the elastic force on the trailing head in rigor state (Fig. 6$a$ or 6$c$), $k_c^{(0)}$ is the ATP-turnover rate under $F_{load} = 0$, and $\delta_c$ is a characteristic distance. $A_c$ is a dimensionless constant that determines the degree to which either mechanical or biochemical events limit the enzymatic cycle at vanishing load: biochemical transitions are rate-limiting for $A_c \ll 1$, whereas mechanical transitions become limiting when $A_c \geq 1$ [43]. Taking $k_c^{(0)} = 30$ s$^{-1}$ [29], $\delta_c = 24$ nm and $F_0 = 0.5$ pN, from Eqs. (3) and (4) we calculate $V$ versus $F_{load}$ for different values of $A_c$. The results are shown in Fig. 9. For other values of $F_0$, we can obtain the same results by adjusting $A_c$. The predicted results in Fig. 9 can be tested in future experiments.

### 3.3. Stall force

Besides the load-dependent step sizes, another distinguishing feature of the cytoplasmic dynein is that the mean stall force, $F_{stall}$, is dependent on the ATP concentration [12]: The mean stall force $F_{stall}$ increases linearly with the increase of [ATP] up to a maximum value at saturating [ATP]. Using our model, we explain this as follows. As noted from Fig. 6$a$, during the relative rotation between the stalk and the stem, if the binding force of the stalk tip to MT is always larger than the magnitude of the force acted on the trailing head, $|F_{trail}| = F_0 - F_{load}/2$, the cytoplasmic dynein becomes stalled. Thus we have for the stall force $F_b(\theta) = F_0 - F_{stall}/2$, i.e.,

$$F_{stall} = 2[F_0 - F_b(\theta)], \qquad (5)$$

where $F_b(\theta)$ is the MT binding force of the stalk tip at the end of rotation of the stalk, as shown in Fig. 6$a$ (dashed lines). As we know, the binding force $F_b(\theta)$ decreases with increase of the rotation angle $\theta$ of the stalk relative to the vertical line to MT. On the other hand, it is experimentally shown that there are four ATP-binding loops, AAA1-AAA4, in each globular domain. The mean number of loops to which ATP binds should depend on [ATP]. At low [ATP], the mean number should increase with the increase of [ATP] and, at high [ATP], all four loops have ATP bound to them. As we assumed before [Point (i) in Section 2], ATP binding to a loop between two AAA modules tightens the contact between them, thus reducing the diameter of the globular domain. The more loops to which ATP binds, the smaller the diameter is, implying that, following product release from one ATPase loop, the larger the rotation angle $\theta$ is. Thus $F_b(\theta)$ decreases with the increase of [ATP] at low [ATP] and becomes saturated at high [ATP]. Accordingly, from Eq. (5) we deduce that $F_{stall}$ increases with the increase of [ATP] at low [ATP] and becomes saturated at high [ATP]. This feature is different from other cargo-transporting dimeric molecular motors such as conventional kinesin, myosin-V and myosin-VI, where the stall forces are almost independent of the ATP concentration.

In addition, since a cytoplasmic dynein has two long stems we thus anticipated that the internal elastic force would be much smaller than that of a kinesin dimer which has two short neck linkers. Therefore, the maximum stall force of dynein, which is about 1 pN [12], is much smaller than that of kinesin, which is about 6 pN [44].

### 3.4. Wobbling behavior

Because the equilibrium distance $d_{equi}^{(0)}$ between the two stalk tips of a cytoplasmic dynein is larger than the period $d$ of MT, it is more probable for the detached leading stalk tip, as shown in Fig. 6b, to bind sites on neighboring MT protofilaments. This is different from kinesin for which the equilibrium distance between the two heads $d_{equi} \approx 5$ nm is smaller than $d = 8$ nm. Thus it is less probable for the detached leading head of a kinesin to bind sites on the neighbouring MT protofilaments that are more than $d = 8$ nm away from the attached trailing head. Therefore, the cytoplasmic dynein exhibits greater lateral movements among MT protofilaments compared with kinesin [45], by which dynein may give way to kinesin when they encounter.

## 4. DISCUSSION

In the conventional models for dynein, it was assumed that the information for a nucleotide-state change located in the globular domain is transited to the MT-binding site in the stalk tip through the 15-nm length of the stalk. But the remaining substantial question of how this transition is communicated is very puzzling. On the contrary, in our present model there is no such an enigmatic communication: Though the dissociation of dynein from MT depends on the nucleotides present in the solution, the binding affinity of stalk tip for MT is independent of the nucleotide state in the globular domain. The change from strong to weak binding is determined naturally by the varying relative orientation between the two interacting surfaces as the stalk rotates following ATP binding or product release. As will be discussed in detail below, this large conformational change of dynein and thus its dissociation from MT is dependent on nucleotides in the solution.

### 4.1. Nucleotide-independent binding affinity of the stalk tip for MT

In our model, either ATP (ADP.Vi) or nucleotide-free (apo) dynein can bind strongly to MT. This is consistent with the experimental result by Spungin et al. [46], where it was demonstrated that, in the presence of a non-hydrolyzable analog of ATP (AMP-PCP), the axonemal dynein, i.e., the Dynein.AMP-PCP binary complex can bind strongly to MT just the same as the nucleotide-free dynein. However, in conventional models, the Dynein.ATP binary complex should bind weakly to MT, which is at odds with this experimental result [46].

Another experiment [21] showed that the dissociation percentage of dynein from MT in the presence of 5 mM AMP-PNP (a non-hydrolyzable analog of ATP) is smaller (24%) than that in the presence of 5 mM ATP (77%). This is consistent with the induction from our model that the binding affinity of dynein for MT is nucleotide independent and only ATP binding/dissociation or release of the ATP hydrolysis product ADP induces the rotation of the stalk, thus inducing the dissociation of dynein from MT. We explain this as follows: In the presence of AMP-PNP, its binding to and slow-rate dissociation from dynein can induce the rotations of the stalk and thus the dissociation of dynein from MT.

In the presence of ATP, its binding to and the release of its hydrolysis product ADP from dynein also induce the dissociation of dynein from MT. Since the dissociation rate of AMP-PNP from dynein is smaller than the ATPase rate, the dissociation percentage of dynein from MT in the presence of AMP-PNP is smaller than that in the presence of ATP. However, according to the conventional models, dynein binds to MT weakly only in ATP and ADP.Pi states and thus the weak binding time for the case of non-hydrolyzable AMP-PNP should be much longer than that for the case of ATP because ADP release is rate limiting in an ATPase cycle [47]. As a result, the dissociation percentage of dynein from MT for the case of AMP-PNP would be larger than that for the case of ATP, which is inconsistent with the experimental results [21].

Note that the experiments by Shpetner et al. [21] also showed that the dissociation percentage in the presence of 5 mM AMP-PNP (24%) is larger than that in the absence of nucleotide (4%). This can be explained as follows: In the case of no nucleotide the dissociation is only resulted from the thermal noise; while in the presence of AMP-PNP, the binding to and dissociation from dynein of AMP-PNP also induce the dissociation of dynein from MT as just discussed above.

A further support to the nucleotide-independent MT binding is the experimental results that the stalk structure itself with inclusion of the tip binds strongly to MT [48,49].

### 4.2. Effects of ATP and ADP bindings on dissociation of dynein from MT

As stated above, according to our model, either ATP binding or product release can induce the dissociation of axonemal dynein from MT due to the rotation of the stalk, rather than that only ATP binding can induce the dissociation. As will be seen below, this argument is consistent with previous experimental results [19,20,22].

The experiment by Shimizu and Johnson [20] showed that when the preincubated MT-dynein was mixed with ADP an incomplete dissociation of dynein from MT with a very slow rate was measured, while when mixed with ATP the dynein dissociated from MT quickly. In the experiment by Holzbaur and Johnson [22], the MT-dynein complex was preincubated with varying concentrations of ADP and then mixed in the stopped flow with varying concentrations of ATP. The results showed that the dissociation kinetics of dynein from MT followed a single exponential at all concentrations of ADP and ATP. The dissociation rate decreases with the increase of [ADP] at fixed [ATP], while increases with the increase of [ATP] at fixed [ADP]. In the following we explain these experimental results using our model.

First, an important result to note in the experiment of Holzbaur and Johnson [22] is that the release rate of ADP from the Dynein.ADP complex is ~1000 $s^{-1}$, where the ADP in the binary complex comes from binding of ADP in the solution. This value is much different from the measured value of ~8–30 $s^{-1}$ in other experiments [29,35], where ADP in the binary complex is produced from the hydrolysis of ATP. Thus there are (at least) two states of the binary complex with ADP, one is denoted by Dynein*.ADP and the other by Dynein.ADP, similar to the case of myosin [50–53]. The pathway for the ADP release is thus

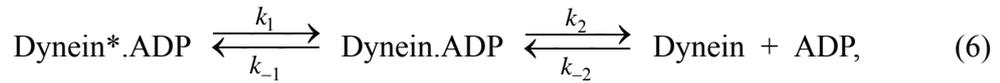

$$\text{Dynein*.ADP} \underset{k_{-1}}{\overset{k_1}{\rightleftarrows}} \text{Dynein.ADP} \underset{k_{-2}}{\overset{k_2}{\rightleftarrows}} \text{Dynein} + \text{ADP}, \quad (6)$$

where the first step is rate limiting with $k_1 \approx$ 8–30 $s^{-1}$, and the ADP binding/release in the experiment by Holzbaur and Johnson [22] corresponds to the second step with $k_2 \approx$ 1000 $s^{-1}$.

Experimental results by Tani and Kamimura [54] determined that Dynein.ADP is a force-generating intermediate. We thus assume, as in Kon et al. [27] and Höök et al. [28], that the transition from Dynein*.ADP to Dynein.ADP involves the conformational change (i.e., power stroke) and the transition from Dynein.ADP to Dynein + ADP involves no (or negligible) conformational change, much similar to the case of myosin [50–53]. Furthermore, because the former transition that is associated with power stroke is accompanied with a large free energy drop, $k_{-1}$ should be very small. Therefore, when the preincubated nucleotide-free MT.dynein is mixed with ADP, because of the very small $k_{-1}$, only a few portion of dyneins involve conformational change with a very slow rate and thus the dissociation of dynein from MT will be incomplete and the dissociation rate is very slow, which is consistent with the experimental result [20]. However, when the preincubated nucleotide-free MT.dynein is mixed with ATP, the rapid ATP binding induces the conformational change and thus dynein dissociates from MT quickly.

When the preincubated MT.Dynein.ADP ternary complex is mixed with ATP, as in experiment by Holzbaur and Johnson [22], the pathway should be

$$\text{Dynein.ADP} \underset{k_{-2}}{\overset{k_2}{\rightleftarrows}} \text{Dynein} + \text{ADP} \overset{k_3}{\longrightarrow} \text{Dynein.ATP} + \text{ADP}, \qquad (7)$$

where, for simplicity, we have neglected the transitions from Dynein.ADP to Dynein*.ADP and from Dynein.ATP to Dynein because of very small $k_{-1}$ and $k_{-3}$. Since the second transition in Scheme (7) involves the conformational change while the first transition involves no conformational change as just discussed above, from Scheme (7) the dissociation rate of dynein from MT due to the second transition is obtained as

$$K = \frac{k_2 k_T [\text{ATP}]}{k_2 + k_D [\text{ADP}] + k_T [\text{ATP}]}, \qquad (8)$$

where $k_{-2} = k_D[\text{ADP}]$ and $k_3 = k_T[\text{ATP}]$ have been used. Thus the experimental results [22] (i.e., at fixed [ATP] the dissociation rate $K$ decreases with increasing [ADP], while at fixed [ADP] $K$ increases with increasing [ATP] and then becomes saturated) can be readily explained. In particular, because the ADP-release rate, $k_2 \approx 1000$ s$^{-1}$, is much higher than the ATP-binding rate in the dissociation experiment [22], the dissociation kinetics follows a single exponential [22]. However, if there is only one state of the ternary MT.Dynein.ADP complex and only ATP binding induced the dissociation from MT, as assumed in the conventional models, the dissociation kinetics would be biphasic, which was never observed in the experiment [22].

**4.3. Activation of product ADP release by MT**

As we mentioned in Point (ii) of Section 2, the strong MT-binding of the stalk tip promotes a conformational change in the active site, which activates the release of the ATP hydrolysis product ADP. This experimental observation is usually understood as resulting from a communication from the MT-binding tip to the globular domain through the stalk, i.e., some conformation change occurs at the contact between the stalk and the globular domain, which is induced by MT binding of the stalk tip. Here we explain this activation as follows. When the stalk tip is in the strong MT-binding state, the tip and stalk remain in a fixed orientation relative to MT [49], similar to the cases of kinesin head binding strongly and sterospecifically to MT [55] and myosin head binding strongly and sterospecifically to actin filament [56]. This fixed orientation of the stalk can result in an internal force (or torque) at the contact between the stalk and the globular domain due to the relative movement between the globular domain and MT driven by the thermal noise. This force (or torque) can thus promote a conformational change in the active site located within the globular domain, accelerating the product release.

**4.4. Effects of ATP and ADP bindings to loops P2 and P4 on MT gliding velocity**

Similar to the effect of ATP binding to loop P1 on ATPase activity of loop P3 or ATP binding to P3 on ATPase activity of P1, as discussed in Subsec. 2.1, ATP binding to loop P2 and/or P4 can also induce a force on loops P1 and P3, thus reducing the ATPase activity of P1 and P3. However, in the presence of ADP, the situation will be different: ADP can bind to P2 and/or P4 instead of ATP and, as in Scheme (7), this ADP.dynein complex has the same conformation as apo-dynein (i.e., an open conformation of loop P2 and/or P4). Thus, as discussed in Subsection 2.1, the ATPase activity of loop P1 and P3 should be larger than that in the case of ATP.dynein complex (i.e., ATP binding to P2 and/or P4 in the absence of ADP). Therefore, ADP can increase the velocity of MT driven by *Chlamydomonas* inner-arm axonemal dyneins, which are in agreement previous experimental results [57–59].

**4.5. Effect of mutation on MT binding affinity**

In the conventional models for dynein movement, the MT binding affinity of the stalk tip is assumed to be regulated by the nucleotide state of the globular domain. That is, some conformational change should take place along the long coiled coil of the stalk, which induces the variation of the MT binding affinity. To study the effect of the alignment between the hydrophobic heptad repeats in the two strands of the stalk coiled coil on MT binding of the stalk tip, Gibbons et al. [60] have recently made a series of chimeric constructs by fusing a short fragment of the stalk (the tip together with 12–36 residues of the stalk) onto a stable coiled coil provided by SRS. It was found that different contructs have different MT-binding affinities depending on the degree of the misalignment between the hydrophobic heptad repeats in the two strands of the coiled coil adjacent to the MT-binding tip.

These interesting results mean that the MT-binding affinity of the stalk tip can indeed be modulated by the conformation of the coiled coil in the region of the stalk adjacent to the tip. Thus it seems reasonable to think that the nucleotide state of dynein can determine the MT-binding affinity of the stalk tip by generating sliding movements between the two strands of the coiled coil structure. However, it should be noted that the length of the stalk coiled coil in these chimeric constructs are only about one fifth of that in intact dynein. In the case of the long stalk of an intact dynein, it is less possible for a misalignment at one end of the stalk to propagate over its full length to the other end, thus greatly reducing the possibility of modulating the MT-binding affinity of the stalk tip by nucleotide changes at the ATPase sites in the globular domain.

It was there further examined that these different MT-binding affinities are the result of local conformational changes in the vicinity of the MT-binding site rather than being due to a global disruption of the MT-binding domain structure [60]. In other words, the local conformational change in the vicinity of the MT-binding site gives a worse interaction between the MT-binding domain and MT. One possible reason for this worse interaction is that the local conformational change in the vicinity of the MT-binding site reduces the contacting surface of the MT-binding site for MT, which has the same effect as the change of the relative orientation between the MT-binding site and MT following the stalk rotation. Another possible reason is that the local conformational change in the vicinity of the MT-binding site induces the change of the charge distribution on the binding site or the hydrophilic property of the binding site, thus resulting in the change of interaction force. If it is the latter case, in our model, the relative orientation change due to the rotation of the stalk for a dynein with such a mutated coiled coil adjacent to the MT-binding domain would induce much weaker interaction compared with native intact dynein.

## 5. CONCLUSIONS

We proposed a model for unidirectional movement of both axonemal and cytoplasmic dyneins based on previous structural observations. In this model it was assumed that the binding affinity of the stalk tip for MT is independent of the nucleotide state in the globular domain. Nevertheless, this assumption is not in contradiction with previous available biochemical experimental results that the dissociation of dynein from MT depends sensitively on the nucleotides present in the solution. Using the model we can explain well the processive movement of MT driven by a single-headed axonemal dynein at saturating ATP concentration. Various experimental results on single cytoplasmic dyneins, such as the step size being an integer times of the period of the MT lattice, the dependence of step size on load, the dependence of stall force on ATP concentration, and the wobbling behaviours, are well explained.

This work was supported by the National Natural Science Foundation of China.

# Figure legends

**Fig. 1.** Proposed model for the stalk rotation relative to the stem of a dynein monomer induced by transition of nucleotide states. The six AAA modules (1–6) and C-terminal sequence (C) are indicated. (a) In ADP.Vi state (i.e., ATP or ADP.Pi state), the globular domain (head) of dynein is compact due to tightening of the contacts between AAA1 and AAA2 and between AAA3 and AAA4 modules. (b) In apo-state, the globular domain is less compact due to loosening of the contacts. The orientation of the stalk relative to the stem becomes different from the case in (a).

**Fig. 2.** Model for the stalk rotation relative to the stem of a dynein monomer induced by transition of nucleotide states as proposed by Burgess et al. [16]. The orientations of AAA1-AAA4 modules relative to each other change as a function of nucleotide binding or release. (a) ADP.Vi state. (b) Apo state.

**Fig. 3.** Microtubule sliding by a fixed axonemal dynein monomer at saturating ATP concentration. Strong binding of the stalk tip occurs when the stalk is perpendicular to MT that is schematically shown in gray. Dashed arrows indicate the moving direction of MT. Here and in the following figures, for simplicity, the globular domain is represented by an orange circle. (a) Dynein binds strongly to MT in ADP.Vi state. (b) Activated by MT, Pi and ADP are rapidly released. The dynein changes to the apo-state conformation, driving MT moving toward the plus end by a distance of $L_S$. Due to the change in the orientation of the stalk relative to MT, the contacting surface between the stalk tip and MT is reduced and the MT-binding strength of the stalk tip decreases. (c) Upon ATP binding, the stalk rotates back and the dynein returns to its original conformational state, binding strongly to MT again. From (a) to (c), an ATPase cycle is finished, with the MT

moving a distance of *nd* (*n* = 1). (a') Dynein binds strongly to MT in apo state. (b') Upon ATP binding, the stalk rotates from the apo-state orientation to the ADP.Vi-state orientation, driving the MT moving toward the minus end. Due to the change in the orientation of the stalk relative to MT, the interaction surface between the stalk tip and MT is reduced and the MT-binding strength of the stalk tip decreases. (c') Due to the very low rate of product release of the dynein without MT activation, within the long-time period of product release, the dynein rebinds strongly to MT while driving MT toward the plus direction or is detached away from MT (not shown). The conformational state is the same as that shown in (a).

**Fig. 4.** Schematic illustration to show the effect of bending elasticity of the stem. The conformation of the dynein-MT system is resulted from the release of ATP-hydrolysis products in Fig. 3(a). Here it is assumed that the stem has some bending elasticity and MT is fixed.

**Fig. 5.** Equilibrium conformations of a cytoplasmic dynein dimer in different nucleotide states. (a) Both heads in ADP.Vi states. (b) One head in ADP.Vi state and the other one in apo-state.

**Fig. 6.** Schematic illustration of an ATPase cycle with effective mechanochemical coupling for a cytoplasmic dynein dimer at saturating ATP. (a) Both heads bind to MT in ADP.Vi states. The products are released earlier from the trailing head. Driven by the internal elastic force and torque, the change in the relative orientation between the stalk and the stem of the trailing head leads to its movement toward the right (dashed lines), thus decreasing the binding force of the stalk tip to MT. (b) After detachment, the trailing head continues to move to its equilibrium position (dashed lines for the stalk). After ATP binding to the new leading head its stalk rotates back (solid lines for the stalk). $d_{equi}^{(0)}$ is the equilibrium distance between the two stalk tips along MT. (c) The stalk of the new leading head binds strongly to MT in the fixed orientation. From (a) to (c) a mechanical step is made.

**Fig. 7.** Schematic illustration of an ATPase cycle with futile mechanochemical coupling for a cytoplasmic dynein dimer at saturating ATP. (a) Both heads bind to MT in ADP.Vi states. (b) The products are released earlier from the leading head. Driven by the internal elastic force and torque, the change in the relative orientation between the stalk and the stem of the leading head makes the dynein dimer have the conformation with a reduced internal force. (c) After ATP binding, the dynein conformation is resumed. From (a) to (c) a futile mechanochemical cycle is completed.

**Fig. 8.** Effects of a backward load on the step size. (a) Schematic illustration of the backward-load-induced decrease of the distance between the two stalk tips along MT before binding of the leading (right) stalk tip to MT. (b) Calculated step-size distribution versus load at saturating ATP by taking $w = 5.5$ nm. (c) Calculated step-size distribution versus load at low ATP concentration. The probability is proportional to the area of the circles.

**Fig. 9.** Calculated mean velocity of a cytoplasmic dynein versus load using Eqs. (3) and (4) with $k_c^{(0)} = 30$ s$^{-1}$, $\delta_c = 24$ nm and $F_0 = 0.5$ pN.

**Figures**

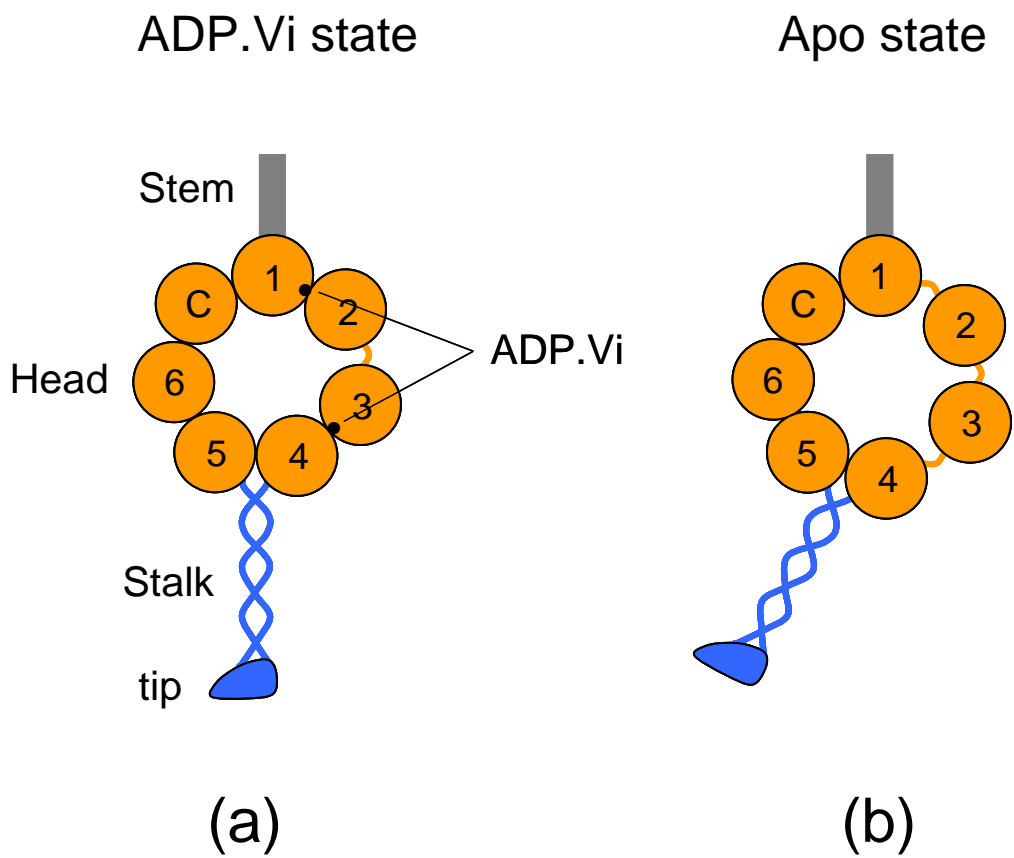

**Fig. 1**

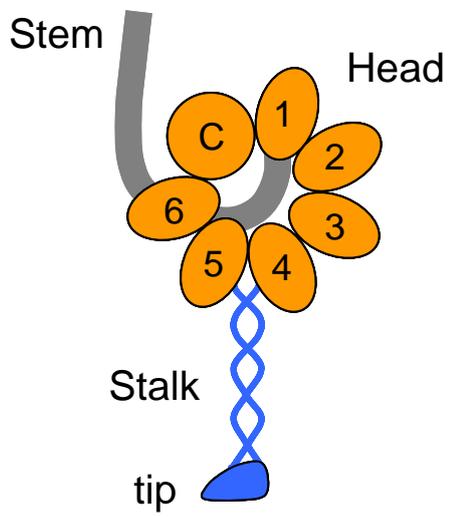
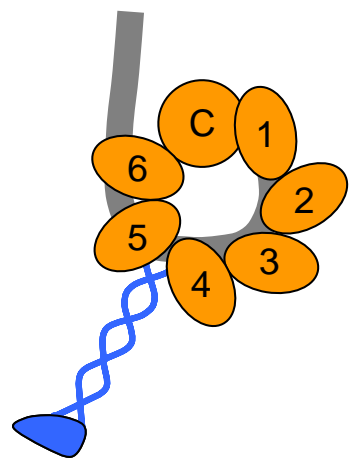

**Fig. 2**

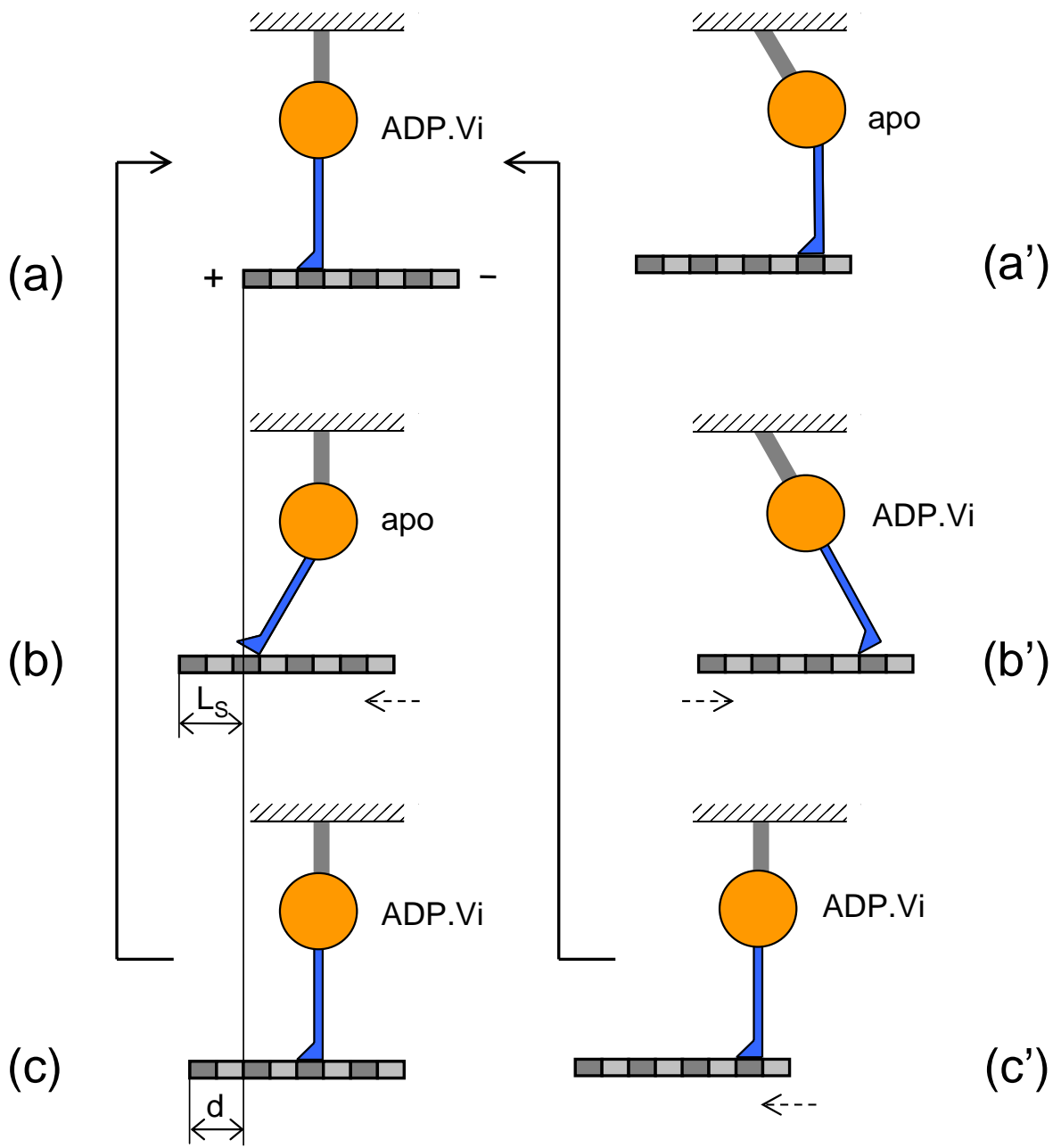

**Fig. 3**

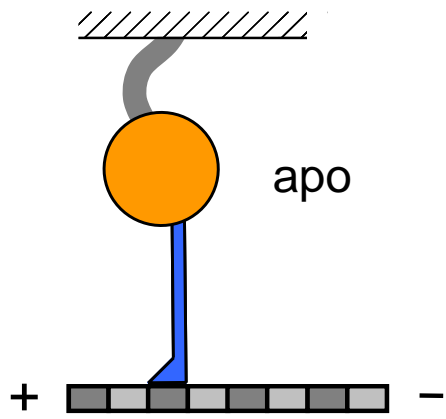

**Fig. 4**

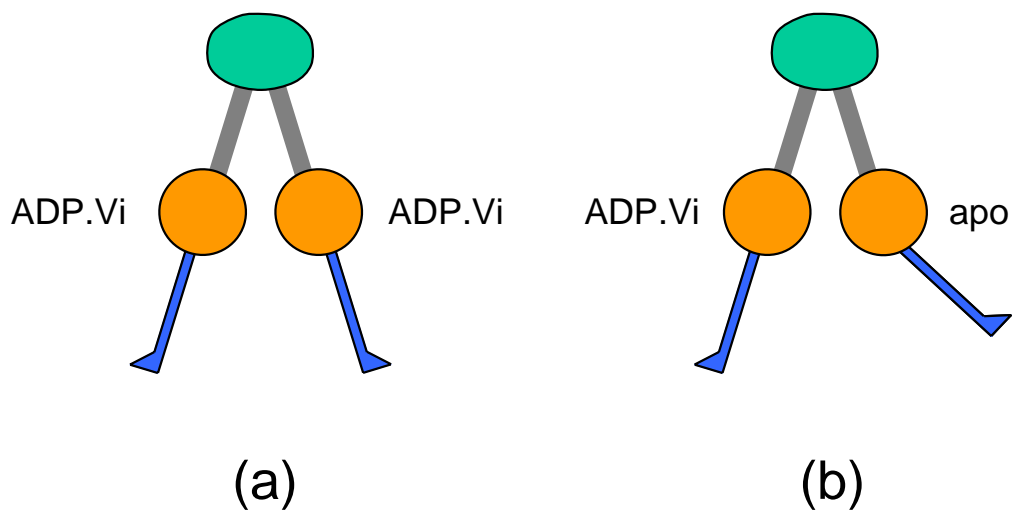

(a)     (b)

**Fig. 5**

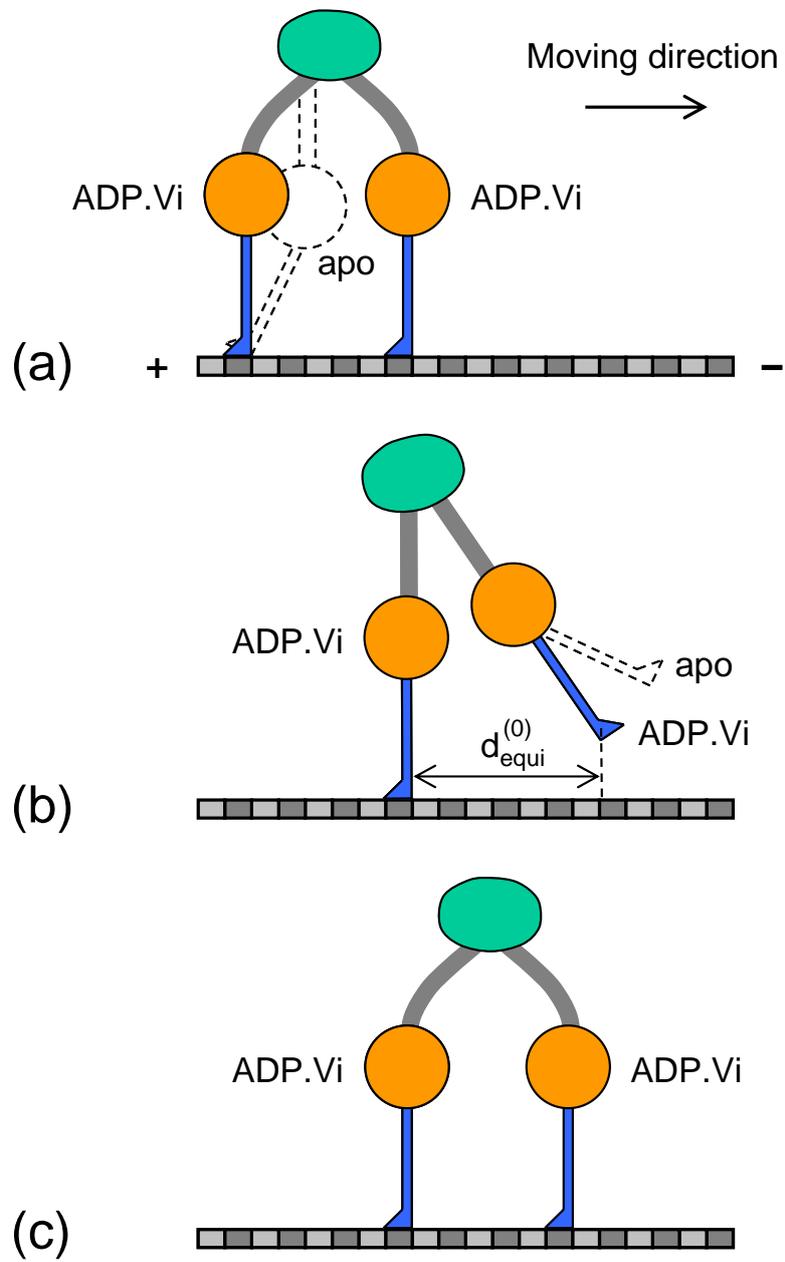

**Fig. 6**

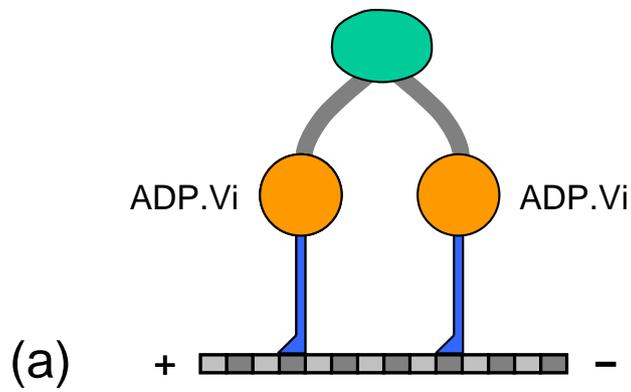

(a)  +  −

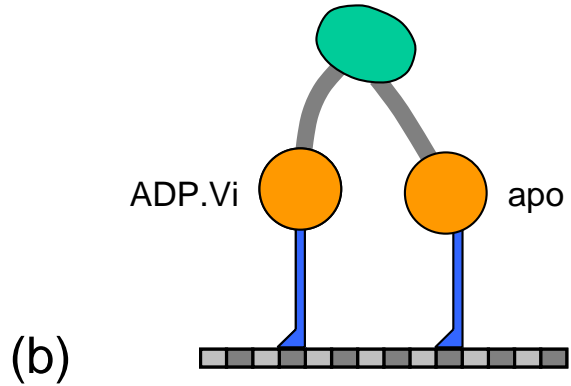

(b)

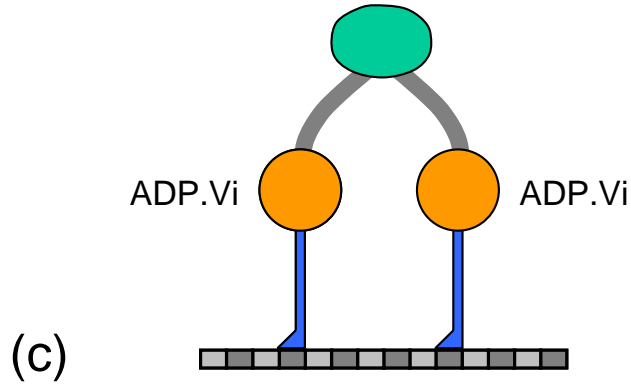

(c)

**Fig. 7**

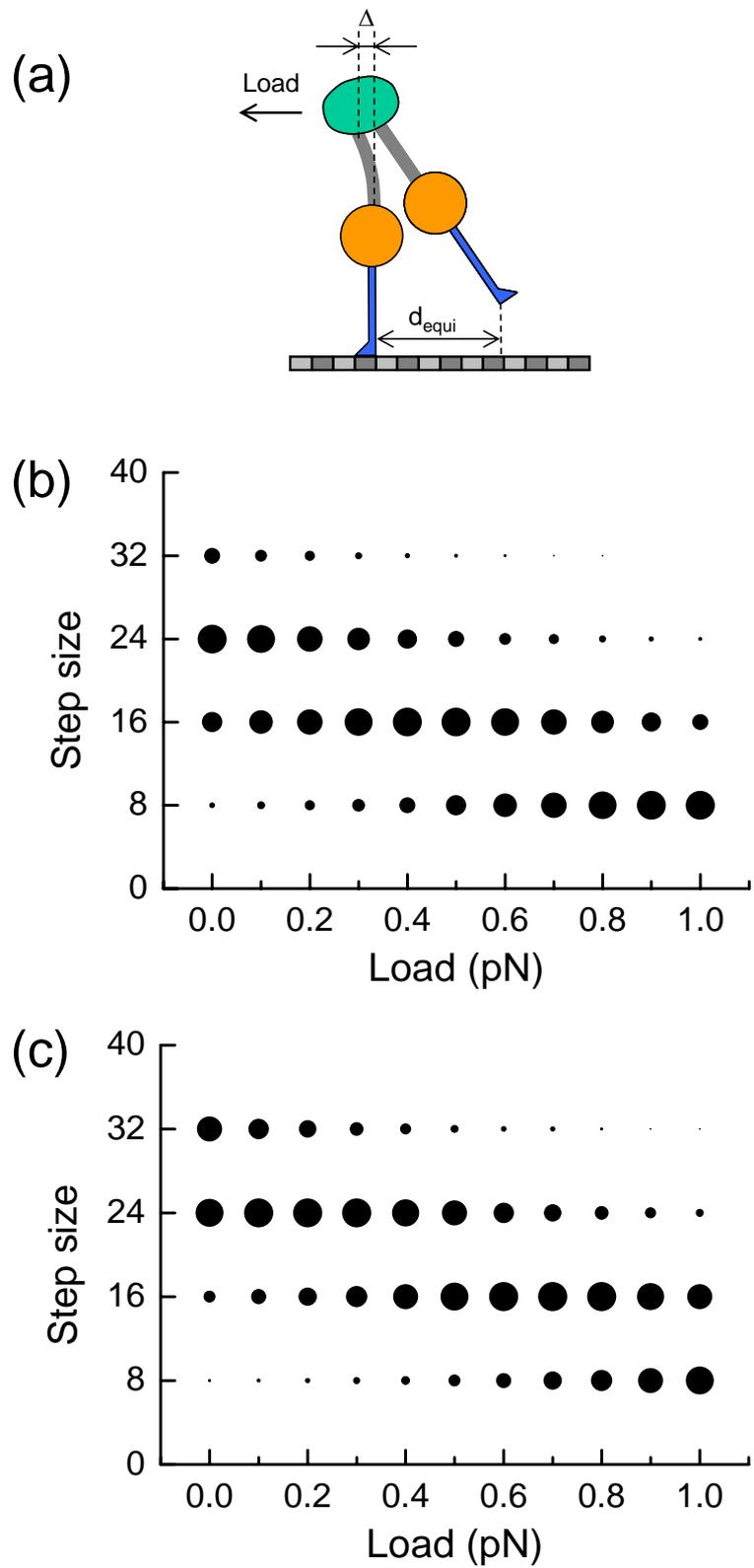

**Fig. 8**

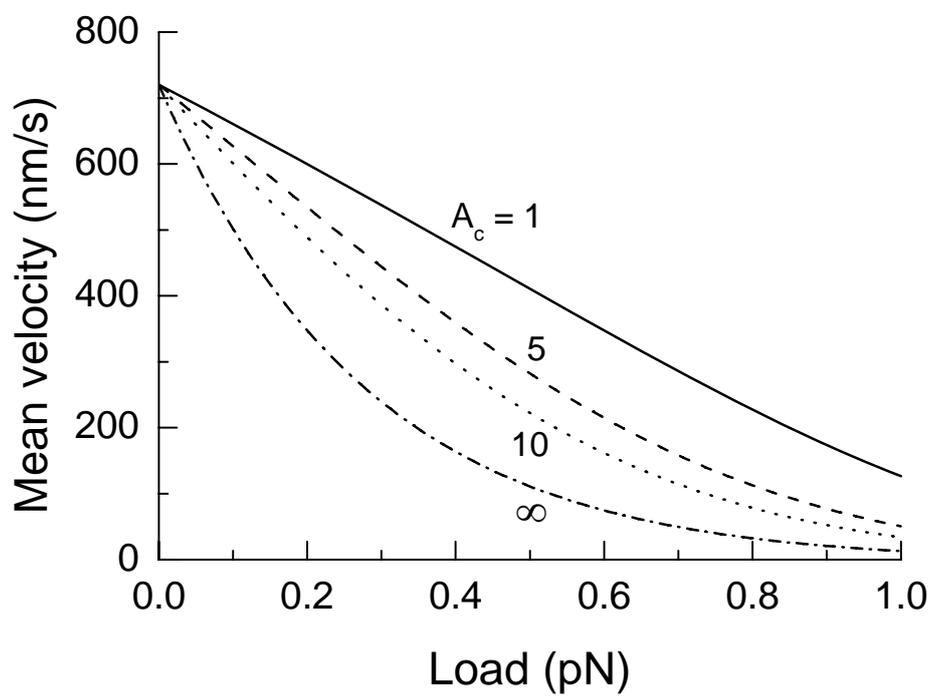

**Fig. 9**